\documentclass{PoS}
\usepackage{epsfig}
\PoS{PoS(LAT2005)181}
\title{Gauge invariance in a $Z_2$ hamiltonian lattice gauge theory }
\ShortTitle{Gauge invariance in a $Z_2$ hamiltonian lattice gauge theory }
\author{\speaker{Takanori Sugihara}\\
  RIKEN BNL Research Center,\\
  Brookhaven National Laboratory,\\
  Upton, New York 11973, USA\\
  E-mail: \email{sugihara@bnl.gov}}
\abstract{
We propose an efficient variational method for $Z_2$ lattice gauge theory 
based on the matrix product ansatz. 
The method is applied to ladder and square lattices. 
The Gauss law needs to be imposed on quantum states to guarantee 
gauge invariance when one studies gauge theory in hamiltonian formalism. 
On the ladder lattice, we identify gauge invariant low-lying states 
by evaluating expectation values of the Gauss law operator 
after numerical diagonalization of the gauge hamiltonian. 
On the square lattice, the second order phase transition 
is well reproduced. 
}

\FullConference{XXIIIrd International Symposium on Lattice Field Theory\\

		25-30 July 2005\\

		Trinity College, Dublin, Ireland}

\begin{document}

\section{Introduction}
The importance of the first-principle study 
in quantum chromodynamics is increasing largely 
because RHIC experiment has started and LHC is also coming. 
For precise description of high-energy heavy ion collisions, 
gauge theory needs to be studied at finite temperature and density 
in a systematic way. 
Ideally, we should also have a methodology for tracing time-evolution 
of quantum states based on the Schr\"odinger equation 
because heavy ion collisions should be treated as 
non-equilibrium evolving systems rather than static. 
Lattice gauge theory is the most useful method for 
studying the quark-gluon systems at zero and finite temperature. 
However, Monte Carlo integration does not work for 
lattice gauge theory with large chemical potential 
because of the severe sign problem. 
It would be worthwhile to pursue a systematic variational approach 
to gauge theory. 
In the previous works, the matrix product ansatz has been 
applied to hamiltonian lattice gauge theory 
on a spatial ladder lattice \cite{Sugihara:2004gx,Sugihara:2005cf}. 

The matrix product ansatz \cite{or} is a simplified version of
DMRG (density matrix renormalization group) 
\cite{white1,Sugihara:2004qr}. 
DMRG has been developed as the method that gives the most 
accurate results for spin and fermion chain models 
such as one-dimensional quantum Heisenberg and Hubbard models 
at zero and finite temperature \cite{ft}.
\footnote{By ``$d$-dimensional'', we mean ($1+d$)-dimensional spacetime.}
DMRG is also useful for 
diagonalization of transfer matrices in two-dimensional classical 
statistical systems \cite{nishino}. DMRG has been extended to 
two-dimensional quantum systems \cite{xiang} and can work 
for bosonic degrees of freedom \cite{Sugihara:2004qr}.

Lattice gauge hamiltonian is obtained by choosing temporal gauge 
in partition function of Euclidean lattice gauge theory. 
In hamiltonian formalism, gauge invariance needs to be 
maintained explicitly by imposing the Gauss law on the Hilbert space. 
On the other hand, Euclidean lattice gauge theory can keep 
gauge invariance manifestly by construction.  
This is one of the reasons why hamiltonian version of lattice 
gauge theory is not popular. In addition, no systematic 
methods had been known for diagonalization of gauge hamiltonian 
before the matrix product ansatz was applied to 
lattice gauge theory in ref. \cite{Sugihara:2004gx}. 
If trial wavefunction is constrained directly with the Gauss law, 
the advantage of the matrix product ansatz is completely spoiled 
because calculation of energy function becomes impossible 
in a practical sense. 
If the hamiltonian is diagonalized without the Gauss law, 
all possible states are obtained including gauge variant states. 
However, it must be possible to extract gauge invariant states 
because all eigenstates of the hamiltonian can be classified 
using generators of the considered gauge group. 
Therefore, if the matrix product ansatz is used, 
we better start from the whole Hilbert space and then 
identify gauge invariant states using the Gauss law operator 
after all calculations. 

\section{Quantum hamiltonian in the $Z_2$ lattice gauge theory}
\label{z2gauge}
We are interested in quantum hamiltonian of the 
$Z_2$ lattice gauge theory. 
Statistical mechanics and quantum hamiltonian are connected 
through the transfer matrix formalism. 
The quantum hamiltonian is obtained by choosing temporal gauge 
in the partition function \cite{Kogut:1979wt} 
\begin{equation}
  H =
  -\sum_{n,i} \sigma_x(n,i)
  -\lambda \sum_{n,i,j} P(n,i,j), 
  \label{hamiltonian}
\end{equation}
where $\sigma_x$ and $\sigma_z$ are Pauli matrices 
and $P$ is a plaquette operator. 
In eq. (\ref{hamiltonian}), 
the first and second summations are taken on the spatial lattice 
for all possible link and plaquette operators, respectively. 
In general, arbitrary states can be represented as a 
superposition of products of $|\pm\rangle_{n,i}$, 
where $\sigma_z(n,i) |\pm\rangle_{n,i} = \pm |\pm\rangle_{n,i}$.

Let us introduce time-independent operators $G(n)$, each of which 
flips spins on all the links emerging from a site $n$ 
\begin{equation}
  G(n) = \prod_{\pm i}\sigma_x(n,i). 
\end{equation}
The operator $G(n)$ defines local gauge transformation 
$G(n)^{-1}HG(n) = H$. 
In order for physical quantities to be gauge invariant, 
quantum states need to be invariant under gauge transformation 
\begin{equation}
  G(n)|\Psi\rangle = |\Psi\rangle. 
  \label{gauss}
\end{equation}
We need to impose the Gauss law (\ref{gauss}) on 
the wavefunction to keep gauge invariance. 
Otherwise, unphysical states may be obtained 
because gauge invariance is not guaranteed.

\section{Matrix product ansatz on a ladder lattice}
\label{mpa}
Since this work is the first application of the matrix product 
ansatz to $Z_2$ gauge theory, 
we would like to consider a simple model. 
The simplest one is a $Z_2$ hamiltonian lattice gauge theory 
on a spatial ladder lattice (see figure \ref{ladder}). 
We assume periodicity in the horizontal direction 
on the ladder for later convenience. In figure \ref{ladder}, 
periodicity is denoted with the open circles. 
\FIGURE{
 \epsfig{file=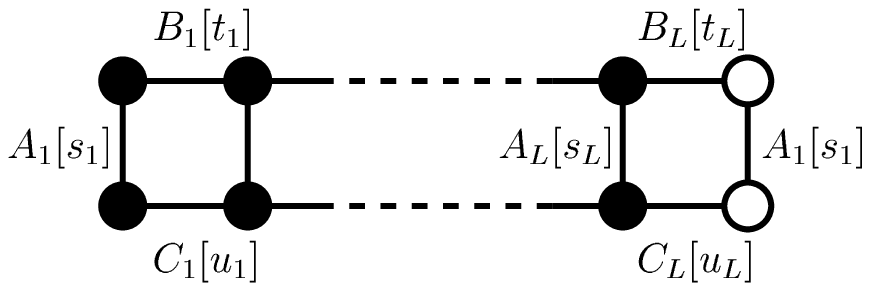,width=8cm}
\caption{A spatial ladder chain with lattice size $L$. 
The open circles indicate periodicity. 
The link variables are dynamical. 
Different sets of matrices are assigned to links. 
}
\label{ladder}
}

The $Z_2$ lattice gauge model has only link variables. 
In our construction, each link is assigned a different set of 
matrices $A_n$, $B_n$, and $C_n$ for parameterization of 
wavefunction (see figure \ref{ladder}). 
The index $n$ represents the $n$-th square on the ladder chain 
and runs from $1$ to $L$. 
The dimension of the matrices is $M$. 
Our matrix product state is give by 
\begin{equation}
  |\Psi\rangle =
  {\rm tr}
  \left(
  \prod_{n=1}^L \sum_{s_n=\pm} \sum_{t_n=\pm} \sum_{u_n=\pm}
  A_n[s_n] B_n[t_n] C_n[u_n]\;
  |s_n\rangle_n |t_n\rangle_n |u_n\rangle_n 
  \right),
  \label{mps}
\end{equation}
where the matrices are multiplied in ascending order keeping 
the order of $A_n B_n C_n$, and the basis states 
$|s\rangle_n$, $|t\rangle_n$, and $|u\rangle_n$ 
are eigenstates of the spin operator $\sigma_z$. 
In this expression, the variables $s$, $t$, and $u$ are used 
to denote the position of the links. 
The implementation of the matrix product ansatz means that 
a ladder lattice has been represented as a one-dimensional system 
with non-nearest neighbor interactions. 
Gauge invariance of matrix product states will be discussed 
in the next section. 

If we require orthogonality of optimum basis states 
according to ref. \cite{or}, we have 
\begin{equation}
  \sum_{j=1}^M \sum_{s=\pm} (X_n[s])_{ij} (X_n[s])_{i'j} =
  \delta_{ii'}, 
  \quad
  \sum_{i=1}^M \sum_{s=\pm} (X_n[s])_{ij} (X_n[s])_{ij'} =
  \delta_{jj'}, 
\end{equation}
where $X$ stands for $A,B$, and $C$. 
If these conditions are not imposed, 
norm of the matrix product state (\ref{mps}) may becomes very small, 
which results in numerical instability. 

Energy 
\begin{equation}
  E = \frac{\langle \Psi|H|\Psi\rangle}{\langle\Psi|\Psi\rangle},
  \label{efunc}
\end{equation}
is a function of the matrices $A_n[s]$, $B_n[t]$, and $C_n[u]$. 
The numerator and denominator can be calculated 
by evaluating trace of a product of $3L$ matrices numerically. 

The minimum of the energy function (\ref{efunc}) corresponds to the 
ground state, which can be obtained based on matrix diagonalization 
as explained below. 
We can reduce the minimization problem (\ref{efunc}) 
into a generalized eigenvalue problem \cite{vpc} 
\begin{equation}
  v^\dagger \bar{H} v = E v^\dagger N v, 
  \label{gen}
\end{equation}
where $\bar{H}$ and $N$ are $2M^2$ by $2M^2$ matrices. 
To understand what is going here, let us consider 
how energy can be minimized by varying $A_n[s]$ 
when other matrices are fixed. 
Note that the hamiltonian and norm matrices are 
bilinear of the matrix $A_n[s]$
\begin{eqnarray}
  \langle \Psi|H|\Psi\rangle &=& \sum_{i,j,k,l} \sum_{s,t}
  (A_n^*[s])_{ij} \bar{H}_{(i,j,s),(k,l,t)} (A_n[t])_{kl}, 
  \\
  \langle \Psi|\Psi\rangle &=& \sum_{i,j,k,l} \sum_{s,t}
  (A_n^*[s])_{ij} N_{(i,j,s),(k,l,t)} (A_n[t])_{kl}, 
\end{eqnarray}
where the matrix $N$ is diagonal for the indices $s$ and $t$. 
Once these expressions are obtained and the variational 
parameters $A_n[s]$ are regarded as a vector $v$, 
the minimization problem (\ref{efunc}) reduces to (\ref{gen}).

\section{Numerical results}
\label{numerical}

The matrix product ansatz assumes large lattice. 
Our lattice size $L=500$ is sufficiently large. 
We solve the generalized eigenvalue problem (\ref{gen}) 
using LAPACK. 
For steady states, real matrices are sufficient for 
parameterizing the matrix product state (\ref{mps}). 
Convergence of energy needs to be checked for the number of 
sweeps and the matrix dimension $M$. 
Energy density $E/L$ converges in accuracy of five digits or higher 
after two sweeps when the matrix size $M$ is fixed. 
\begin{table}
\begin{center}
\begin{tabular}{lllllll}
\hline
$M$ & $E_0/L$ & $E_1/L$ & $E_2/L$ & $E_3/L$ & $E_4/L$ & $E_5/L$\\
\hline\hline
\multicolumn{5}{c}{$\lambda=0.1$} \\
\hline
$2$ &
$\underline{-3.001}$ & $-2.997$ & $-2.997$ &
$-2.997$ & $\underline{-2.993}$ & $\underline{-2.993}$ \\
$3$ &
$\underline{-3.001}$ & $-2.997$ & $-2.997$ &
$-2.997$ & $\underline{-2.994}$ & $\underline{-2.993}$ \\
$4$ &
$\underline{-3.001}$ & $-2.997$ & $-2.997$ &
$-2.997$ & $\underline{-2.997}$ & $\underline{-2.995}$ \\
\hline
\multicolumn{5}{c}{$\lambda=1$} \\
\hline
$2$ &
$\underline{-3.124}$ & $-3.121$ & $-3.121$ &
$-3.118$ & $\underline{-3.114}$ & $\underline{-3.112}$ \\
$3$ &
$\underline{-3.124}$ & $-3.121$ & $-3.121$ &
$-3.118$ & $\underline{-3.114}$ & $\underline{-3.112}$ \\
$4$ &
$\underline{-3.124}$ & $-3.121$ & $-3.121$ &
$-3.118$ & $\underline{-3.114}$ & $\underline{-3.112}$ \\
\hline
\multicolumn{5}{c}{$\lambda=10$} \\
\hline
$2$ &
$\underline{-10.27}$ & $-10.27$ & $-10.27$
& $\underline{-10.27}$ & $-10.23$ & $\underline{-10.23}$ \\
$3$ &
$\underline{-10.27}$ & $-10.27$ & $-10.27$
& $\underline{-10.27}$ & $-10.26$ & $\underline{-10.23}$ \\
$4$ &
$\underline{-10.27}$ & $-10.27$ & $-10.27$
& $\underline{-10.27}$ & $-10.26$ & $\underline{-10.23}$ \\
\hline
\end{tabular}
\end{center}
\caption{Energy density $E/L$ of six low-lying states is listed 
for $\lambda=0.1,1$, and $10$ when lattice size is $L=500$. 
Good convergence of energy is obtained with small $M$. 
}
\label{conv}
\end{table}

\begin{figure}
\begin{center}
\epsfig{file=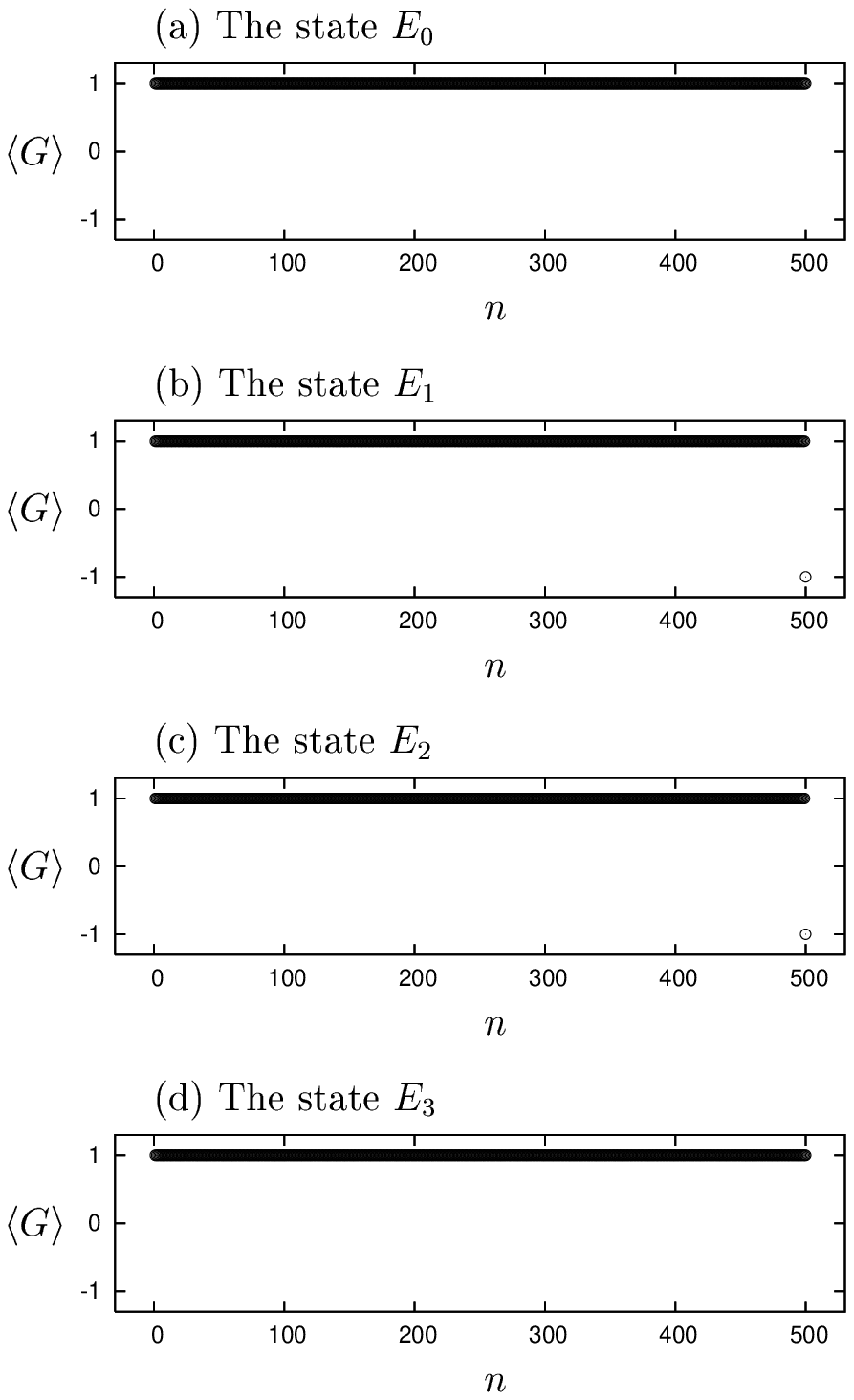, width=.4\textwidth}
\end{center}
\caption{Expectation values of the Gauss law operator is 
plotted for the four low-lying states (a) $E_0$, (b) $E_1$, 
(c) $E_2$, and (d) $E_3$ with $\lambda=10$, $L=500$, and $M=4$. 
The circles are the calculated values. 
The states (a) and (d) are gauge invariant because the Gauss law 
is satisfied on every lattice sites. 
On the other hand, the states (b) and (c) are gauge variant 
because $\langle G(500)\rangle =-1$. 
These statements hold in accuracy of seven digits or higher. 
}
\label{gl}
\end{figure}

Table \ref{conv} shows energy spectra of six low-lying states 
for three values of the coupling constant: $\lambda=0.1,1$, and $10$. 
The sweep process has been repeated twice. 
In this model, convergence of energy is very fast 
in contrast to Heisenberg chains \cite{Sugihara:2004gx}\cite{or}. 
Small matrix dimension is sufficient for good convergence. 
Since we have obtained low-lying states without imposing 
the Gauss law on the variational space, 
gauge variant states are contained. 
In table \ref{conv}, gauge invariant states are denoted 
with underlines. The other states are gauge variant. 
As we will see, gauge invariant physical states can be identified 
by calculating expectation values of the Gauss law operator.

In the ladder chain model, the Gauss law operator $G(n)$ is 
a product of three $\sigma_z$ operators 
(two horizontal and one vertical). 
We evaluate expectation values of $G(n)$ on the upper lattice 
sites shown in figure \ref{ladder}. 
Then, the number of the Gauss law operators to be evaluated is $L$. 
Expectation values on the lower 
sites are same as the upper ones because of reflection symmetry. 
Figures \ref{gl} plots expectation values of the Gauss law operator 
$\langle G(n) \rangle$ in the case of $\lambda=10$ for the states 
(a) $E_0$, (b) $E_1$, (c) $E_2$, and (d) $E_3$. 
In figures \ref{gl} (a) and (d), the Gauss law $G(n)=1$ is satisfied 
uniformly on every lattice sites. Therefore, 
the obtained states $E_0$ and $E_3$ are gauge invariant. 
On the other hand, in figures \ref{gl} (b) and (c), 
the states $E_1$ and $E_2$ are gauge variant 
because gauge symmetry is definitely broken at the site $n=500$. 
The position of this special lattice site depends on 
where the sweep process ends. 
The relation $\langle G(n) \rangle = 1$ or $-1$ holds for 
the obtained low-lying states in accuracy of seven digits or higher 
when $M=4$. 
In this way, we can classify the obtained states into 
gauge invariant states and others. 

\section{Extension to square lattice}
We apply the matrix product ansatz to 
(2+1)-dimensional $Z_2$ lattice gauge theory on a square lattice, 
which  has a second order phase transition. 
It is possible to solve the model in the same way as before 
without imposing the Gauss law on a variational space. 
However, we solve the Gauss law 
analytically to reduce calculation load. 
As a result, the model is equivalent to the transverse field Ising model. 
The square lattice is organized into one-dimensional lattice 
so that the matrix product ansatz can be applied. 
The non-local interactions can be handled 
by increasing the dimension of the matrix size. 
The matrix size used for the calculation is $M=30$. 
The obtained value of the critical coupling is 
$\lambda_{\rm c} \sim 3.12$, 
which is close to the past numerical results. 
However, our lattice size $L=12$ is still small.  
Further refinement will be given elsewhere.

\end{document}